# Ergodicity of non-Hamiltonian Equilibrium Systems

Denis J. Evans[1], Stephen R. Williams[2], Lamberto Rondoni[3,4] and Debra J. Searles[5,6]

[1]Department of Applied Mathematics
Research School of Physics and Engineering
Australian National University
Canberra ACT, 0200
Australia

[2]Research School of Chemistry
Australian National University
Canberra ACT, 0200
Australia

[3]Dipartimento di Scienze Matematiche and Graphene@Polito Lab, Politecnico di Torino,
Corso Duca degli Abruzzi 24, 10129 Torino, Italy

[4]INFN, Sezione di Torino, Via P. Giuria 1, 10125 Torino, Italy

[5]Australian Institute for Bioengineering & Nanotechnology
The University of Queensland, Brisbane QLD, 4072
Australia

[6]School of Chemistry and Molecular Biosciences
The University of Queensland, Brisbane QLD, 4072
Australia

**Abstract**

It is well known that ergodic theory can be used to formally prove a weak form of relaxation to equilibrium for finite, mixing, Hamiltonian systems. In this Letter we extend this proof to any dynamics that preserves a mixing equilibrium distribution. The proof uses an approach similar to that used in umbrella sampling, and demonstrates the need for a form of ergodic consistency of the initial and final distribution. This weak relaxation only applies to averages of physical properties. It says nothing about whether the distribution of states relaxes towards the equilibrium distribution or how long the relaxation of physical averages takes.

**Introduction**

It is known from ergodic theory that a finite, autonomous, Hamiltonian system that preserves a mixing microcanonical equilibrium distribution, will, from almost any initial state described by an phase space distribution $f(\mathbf{\Gamma};0)\delta(H(\mathbf{\Gamma})-E)\delta(\mathbf{P})$, eventually have averages of physical phase functions that relax towards their microcanonical equilibrium values [1]. In the expression above $H(\mathbf{\Gamma})$ is the Hamiltonian, $E$ the energy and $\mathbf{P}$ the total linear momentum of the system.

A system is said to be *mixing* if for integrable, physical phase functions, time correlation functions computed with respect to a stationary distribution factorize into products of averages computed with respect to the same distribution:

$$\lim_{t\to\infty}\left\langle A(\mathbf{\Gamma})B(S^t\mathbf{\Gamma})\right\rangle_\infty - \left\langle A(\mathbf{\Gamma})\right\rangle_\infty \left\langle B(\mathbf{\Gamma})\right\rangle_\infty = 0 \; . \tag{1}$$

Here the brackets, $\left\langle ...\right\rangle_\infty$, denote an ensemble average with respect to an invariant (*i.e.* time-stationary) probability measure $\mu_\infty$. In the case that $\mu_\infty$ has a density $f(\mathbf{\Gamma};\infty)$, one may write:

$$\left\langle A\right\rangle_\infty = \int d\mu_\infty(\mathbf{\Gamma})A(\mathbf{\Gamma}) = \int d\mathbf{\Gamma}\, f(\mathbf{\Gamma};\infty)A(\mathbf{\Gamma}) \tag{2}$$

where $f(\mathbf{\Gamma};\infty) = \left|\partial\mu_\infty/\partial\mathbf{\Gamma}\right|$ is a normalized distribution and $f(\mathbf{\Gamma};\infty)d\mathbf{\Gamma}$ is dimensionless.

If on the other hand $f(\Gamma;\infty)$ is not defined over the phase space of the system, one could write only the first equality $\langle A \rangle_\infty = \int d\mu_\infty(\Gamma) A(\Gamma)$, where $d\mu_\infty(\Gamma)$ is dimensionless and normalized.

Implicit in this definition is the fact that the invariant measure must be preserved by the dynamics, in which case at all times $\langle B(S^t\Gamma) \rangle_\infty = \langle B(\Gamma) \rangle_\infty$ for all phase variables, B. So mixing systems must, as a prerequisite, have an invariant measure that is preserved by the dynamics *and* additionally they must satisfy (1) with respect to this same invariant distribution or measure.

We note that if the system has nonzero angular momentum no stationary long-time measure is possible (unless we transform to a non-inertial, co-rotating coordinate frame where Hamiltonian dynamics breaks down). So if angular momentum is conserved in our system we must set it to zero as is done for the total linear momentum.

The mixing property is a property of the stationary state of interest, in which observables take the average values denoted by $\langle ... \rangle_\infty$. It represents the fact that, in the macroscopically stationary state, correlations among microscopically evolving physical properties (physically relevant phase functions) decay in time. Therefore, in general the mixing condition *would not appear* to guarantee relaxation to an invariant state. Mixing already *assumes* stationarity of the macrostate and its preservation by the system's dynamics regardless of whether it is reached asymptotically in time, as implied by our notation, or it is initially prepared in that state by some means (*e.g.* a Monte Carlo process).

For completeness we repeat here our version [2, 3] of the standard ergodic theory proof of relaxation for autonomous Hamiltonian systems. We begin by noting that the *microcanonical distribution*, $f_{\mu c}(\mathbf{\Gamma})$ for the infinitely thin energy shell

$D(N,V,E,\delta E): E < H(\mathbf{\Gamma}) < E + \delta E$, which we refer to as $D(\delta E)$:

$$f_{\mu c}(\mathbf{\Gamma}) \equiv \begin{cases} \lim_{\delta E \to 0} \dfrac{1}{\int_{\mathbf{\Gamma} \in D(\delta E)} d\mathbf{\Gamma}}, & \mathbf{\Gamma} \in D(\delta E) \\ \\ 0, & \mathbf{\Gamma} \notin D(\delta E) \end{cases} \qquad (3)$$

is a time-stationary distribution, preserved by the autonomous Hamiltonian dynamics. [4] One can see from the phase continuity equation (often referred to as the Liouville equation) that both $df_{\mu c}(\mathbf{\Gamma})/dt = 0$ and $\partial f_{\mu c}(\mathbf{\Gamma})/\partial t = 0$. In (3) the domain, *D*, is the limiting thin shell domain noted in the integral (3). We note that if the dynamics has constant (zero) total momentum one of the particles may be removed from the phase space since its position and momentum is known from the $3(N-1)$ coordinates and momenta in the system. This also removes a source of non-independence in the phase space variables that would otherwise make the evaluation of $(\partial/\partial \mathbf{\Gamma}) \cdot \dot{\mathbf{\Gamma}}(\mathbf{\Gamma}) = 0$ difficult.

We also note that for finite $\delta E$ our system cannot be T-mixing[1] and there are infinitely many distributions that have $\partial f(\mathbf{\Gamma})/\partial t = df(\mathbf{\Gamma})/dt = 0$. However in the limit $\delta E \to 0$ the system may be T-mixing. In this limit the density is equivalent to the energy

---

[1] A weakly T-mixing system [3] is one where $\lim_{t \to \infty} \left[ \langle A(\mathbf{\Gamma}) B(S^t \mathbf{\Gamma}) \rangle_0 - \langle A(\mathbf{\Gamma}) \rangle_0 \langle B(S^t \mathbf{\Gamma}) \rangle_0 \right] = 0$. Here *A* and *B* are phase variables, and the ensemble averages are over the initial measure. This condition differs from the mixing condition, which refers to the invariant measure rather than the initial measure. The strong form of T-mixing ensures that the correlations not only decay, but decay quickly enough so that $\lim_{t \to \infty} \int_0^t \langle A(\mathbf{\Gamma}) B(S^t \mathbf{\Gamma}) \rangle_0 - \langle A(\mathbf{\Gamma}) \rangle_0 \langle B(S^t \mathbf{\Gamma}) \rangle_0 \, ds = L_A \in \mathbf{R}$. A weak form of T-mixing is sufficient for the cases discussed in this paper.

hypersurface areal density, $\sigma(\mathbf{\Gamma})$:

$$\sigma(\mathbf{\Gamma}) = \frac{\|\dot{\mathbf{\Gamma}}\|^{-1}}{\int_{H(\mathbf{\Gamma})=E} d\mathbf{\Gamma} \|\dot{\mathbf{\Gamma}}\|^{-1}}, \quad \forall \mathbf{\Gamma}: H(\mathbf{\Gamma}) = E \qquad (4)$$

and the system is T-mixing. This expression for the density is obtained by realising that for the energy shell $\delta E = \delta \Gamma \|\nabla H\| = \delta \Gamma \|-\mathbf{J}\dot{\mathbf{\Gamma}}\| = \delta \Gamma \|\dot{\mathbf{\Gamma}}\|$ where $\mathbf{J}$ is the usual symplectic matrix $\begin{pmatrix} 0 & 1 \\ -1 & 0 \end{pmatrix}$ and $\delta\Gamma$ is the limiting shell thickness measured using some phase space distance metric. $\delta\Gamma$ is scalar and a function of $\mathbf{\Gamma}$, and

$$\int_{\mathbf{\Gamma} \in D(\delta E)} d\mathbf{\Gamma} = \int_{H(\mathbf{\Gamma})=E} d\mathbf{\Gamma} \, \delta\Gamma(\mathbf{\Gamma}).$$

We will now give the standard proof that if our ensemble is initially *not* distributed according to this distribution, the ensemble will in fact relax towards this distribution - at least for the purposes of computing time averages of low order physical phase functions. Note: low order physical phase functions are functionals of low order (singlet, pair and triplet, for example) distribution functions. [3] The full $N$-particle phase space distribution is a function of ~6($N$-1) variables where $N$ may be of order of Avogadro's Number if the system is macroscopic.

We compute the time dependent average of a physical phase function $A(\mathbf{\Gamma})$ for some smooth distribution function, $f(\mathbf{\Gamma};t)$ that evolves in time according to the Hamiltonian dynamics and which is non-singular in the domain, $D$:

$$\langle A \rangle_t = \int_D d\Gamma A(\Gamma) f(\Gamma;t)$$

$$= \int_D d\Gamma A(S^t\Gamma) f(\Gamma;0), \qquad (5)$$

where the second equality is due to the equivalence of the Heisenberg and Schrödinger representations of phase space averages [5] and the notation $\langle A \rangle_t$ refers to an ensemble average with respect to the time evolved distribution $f(\Gamma;t)$. In (5) stationarity of the distribution $f(\Gamma;t)$ is not assumed (so $f(\Gamma;t)$ is not necessarily equal to $f(\Gamma;0)$), however, since the dynamics is driven by an autonomous Hamiltonian, the energy is fixed.

Now we multiply and divide the last expression in (5) by the (necessarily finite!) ostensible volume of the phase space. This casts the first line in a form to which the mixing property can (formally) be applied,

$$\langle A \rangle_t = \frac{1}{\int_D d\Gamma} \cdot \int_D d\Gamma \, A(S^t\Gamma) f(\Gamma;0) \cdot \int_D d\Gamma$$

$$\equiv \langle A(S^t\Gamma) f(\Gamma;0) \rangle_{\mu c} \cdot \int_D d\Gamma \qquad (6)$$

We emphasise that in order to derive (6) the ostensible phase space volume needs to be finite and non-zero. This equation also requires that the asymptotic phase space density is nonsingular in $D$.

A few more words need to be said about $\langle A(S^t\mathbf{\Gamma})f(\mathbf{\Gamma};0)\rangle_{\mu c}$. This function is an equilibrium microcanonical, cross-time correlation function. It results from the fact that for Hamiltonian dynamics, any time dependent nonequilibrium ensemble average, say $\langle A\rangle_t$ (i.e. the ensemble average computed with respect to the evolved distribution), equals a time dependent nonequilibrium average $\langle A(S^t\mathbf{\Gamma})\rangle_0$ computed with respect to the initial distribution $f(\mathbf{\Gamma};0)$ (see equation (5)). It also assumes that $f(\mathbf{\Gamma};0) \equiv f_0(\mathbf{\Gamma})$ is a phase function (i.e. it is defined at all $\mathbf{\Gamma}$ and its time invariance is indicated by the '0'). In order for this to be possible we assume that the initial distribution is smooth; it could be an equilibrium distribution for a different dynamics. In such cases the **distribution** function behaves like a low order phase function, to which the mixing condition can be applied. However, and very importantly, for autonomous Hamiltonian systems the propagator for a phase function and a distribution function are not the same. They are in fact Hermitian adjoints. [5] For thermostatted systems the relationship between these two propagators is more complex [5] but of course, the Schrodinger/Heisenberg equivalence is preserved.

We exclude the case where the initial distribution is a nonequilibrium steady state distribution. This is because these distributions are always singular.[2]

Using (1) and knowing that the microcanonical distribution is preserved by the autonomous Hamiltonian dynamics, we now take the long time limit and use the mixing assumption (1), to allow us to factorize the naturally invariant (microcanonical) time correlation function $\langle A(S^t\mathbf{\Gamma})f(\mathbf{\Gamma};0)\rangle_{\mu c}$ into a product of two invariant (microcanonical) averages:

$$\lim_{t\to\infty}\langle A\rangle_t = \langle A(\mathbf{\Gamma})\rangle_{\mu c}\langle f(\mathbf{\Gamma};0)\rangle_{\mu c}\cdot\int_D d\mathbf{\Gamma}$$

$$= \langle A(\mathbf{\Gamma})\rangle_{\mu c}\frac{1}{\int_D d\mathbf{\Gamma}}\int_D d\mathbf{\Gamma}\, f(\mathbf{\Gamma};0)\cdot\int_D d\mathbf{\Gamma} \ . \qquad (7)$$

$$= \langle A(\mathbf{\Gamma})\rangle_{\mu c}.1 = \langle A(\mathbf{\Gamma})\rangle_{\mu c}$$

---

[2] To extend to nonequilibrium steady states (NESS), we would need to assume that D is the domain of the NESS, which is singular in the ostensible phase space, as considered by Sinai. [1] However its topology is generally unknown.

To obtain the final line we use the normalization of the initial distribution function. Equation (1) only applies to a stationary state, however we note that we do not need to *assume* the existence of a stationary state, since as noted above, the microcanonical distribution is indeed preserved by Hamiltonian dynamics.

So $\langle A\rangle_t$ tends towards a microcanonical average, whatever physical phase function $A(\mathbf{\Gamma})$, or initial probability density $f(\mathbf{\Gamma};0)$ one considers - as long as it lies in the phase space domain *D*. This amounts to a formal proof of weak relaxation *towards* the microcanonical equilibrium state denoted by $\langle.\rangle_{\mu c}$. The proof only shows the relaxation of averages of physical quantities towards their equilibrium values. It does not in itself prove that the phase space density relaxes towards the equilibrium density. This proof also carries no information about fundamental physical properties such as relaxation times that are fundamental, *e.g.* for the existence of transport coefficients.

Because distribution functions are normalized we see that for any distribution and time,

$$\langle f(\mathbf{\Gamma};t)\rangle_{\mu c} = \lim_{\delta E \to 0} \int_{D(\delta E)} d\mathbf{\Gamma}\, f(\mathbf{\Gamma};t) / \int_{D(\delta E)} d\mathbf{\Gamma}$$

$$= \lim_{\delta E \to 0} 1 / \int_{D(\delta E)} d\mathbf{\Gamma} = \exp[-S_{\mu c}(E,N,V)/k_B]$$

(8)

where we use the Gibbs definition of entropy $S_{\mu c} \equiv -k_B \int_D d\mathbf{\Gamma}\, f_{\mu c}(\mathbf{\Gamma}) \ln[f_{\mu c}(\mathbf{\Gamma})]$ and

$f_{\mu c}(\mathbf{\Gamma}) = \lim_{\delta E \to 0} 1 / \int_{\mathbf{\Gamma} \in D(\delta E)} d\mathbf{\Gamma}$. The microcanonical average of a normalized distribution function tells us *nothing* about that distribution. The average only tells us the Gibbs entropy of the microcanonical distribution used to calculate the average.

Unless one starts at $t=0$ with the microcanonical distribution, the proof given in (5) and (6) shows (formally) that *averages* of thermodynamic quantities that are functionals of low order distribution functions [3] *approach* microcanonical averages in the long time limit. The actual distribution never *becomes* the microcanonical distribution. At any time, no matter how large, we can always apply a time reversal map and return (eventually!) to the initial distribution. As time increases in the relaxation process, the distribution function becomes ever more tightly folded upon itself, never *becoming* the smooth microcanonical equilibrium distribution. However, this finely detailed folding is ultimately unphysical. Empirically we are only interested in averages we can measure in the laboratory; averages of low order physical properties such as stress, energy, pressure or indeed the dissipation function itself. In the Green expansion of the 6N-dimensional phase space distribution, the averages of physical quantities only depend on the low order terms. This is precisely why we can replace the possibly singular high-order distribution function with a smooth approximation such as $e^{h(\mathbf{\Gamma})}$ where $h(\mathbf{\Gamma})$ is a smooth function. This in turn means that we only need to be concerned with relatively smooth phase space distributions which have to arbitrary precision have the same averages of those physical properties.

Relaxation of the phase space distribution to equilibrium cannot take place. If relaxation was to occur in finite time one could never return to the initial distribution by applying a time reversal map, which would be inconsistent with the reversibility of the equations of motion. This also illustrates that relaxation of phase averages does not *necessarily* mean the relaxation of their densities. This present proof only establishes that probability densities only relax in the weak sense corresponding to relaxation of averages of observables. This is one further reason why treating a probability density like a standard phase variable, as *e.g.* in Eq. (6), is physically unjustified, in general. This also illustrates that caution is needed in treating probability densities. The exact classical distribution function $f(\mathbf{\Gamma}(t);t)$ has explicit time dependence and cannot be treated as a phase variable, even though $f(\mathbf{\Gamma}(t);0) \equiv f_0(\mathbf{\Gamma}(t))$ can.

We now generalize this derivation so that it applies to any dynamics that preserves a mixing equilibrium distribution, $f_{eq}(\mathbf{\Gamma})$ for a finite system. The phase space vector could be augmented by additional variables such as the Nosé-Hoover thermostat multiplier or the system volume to cover a variety of different systems (*e.g.* Nosé-Hoover thermostatted dynamics or Nosé-Hoover isothermal isobaric dynamics). However, we do not show this explicitly in our notation.

So far as we are aware, the ergodic theory proof of relaxation to equilibrium, never introduced a complete mathematical definition of an equilibrium distribution. This definition was not given until our T-mixing proofs of relaxation to equilibrium. For instance if $f_{eq}(\mathbf{\Gamma})$ is a solution of the possibly thermostatted and/or barostatted dynamics it is clear that, $\partial f_{eq}(\mathbf{\Gamma})/\partial t = 0$ but it is well known that in general $df_{eq}(\mathbf{\Gamma})/dt \neq 0$. The question of deciding when a system is in equilibrium or not, or indeed whether a specific dynamics is capable of preserving an equilibrium distribution, is in general not completely straightforward.

From our equilibrium relaxation theorems we define an equilibrium system to be a combination of a distribution function $f(\mathbf{\Gamma};t)$ and a dynamics $\dot{\mathbf{\Gamma}}(\mathbf{\Gamma}(t))$ for which the dissipation function is identically zero (almost) everywhere in the accessible phase space domain; $\Omega(\mathbf{\Gamma}) = 0, \quad \mathbf{\Gamma} \in D$. The instantaneous dissipation function is defined as:

$$\Omega(\mathbf{\Gamma}(t)) \equiv -\partial/\partial\mathbf{\Gamma} \bullet \dot{\mathbf{\Gamma}}(t) - \dot{\mathbf{\Gamma}}(\mathbf{\Gamma}(t)) \bullet \partial \ln(f(\mathbf{\Gamma}(t);0))/\partial\mathbf{\Gamma} \qquad (9)$$

For a specified initial phase space distribution the instantaneous dissipation function is just another (very important) phase function. The dissipation theorem confirms that if a system is sampled from an equilibrium distribution of states, that distribution must be preserved by the dynamics for all time.

It may be non-trivial to establish whether a system is in equilibrium or not. For example if we have a system of interacting particles where the wall particles which act as a thermal reservoir are subject to a μ-thermostat where the equation of change for the $\beta^{th}$

Cartesian component particle momentum $\dot{p}_{\beta,i} = F_{\beta,i} - \alpha p_{\beta,i}^{\mu} Sgn(p_{\beta,i})$ where $F_{\beta,i}$ is the $\beta^{th}$ Cartesian component of the standard interatomic force felt by particle $i$, due to all the other particles in the system and α is chosen to fix say, the sum of the $(\mu+1)^{th}$ moments of the particle momenta with a Gaussian thermostat, equilibrium is only possible if $\mu = 1$. This can only be understood using our definition of an equilibrium system. If $\mu \neq 1$ the system is autodissipative and equilibrium is not possible. [8]

In our proof [3,4,6] of relaxation to equilibrium using the ΩT-mixing[3] property, we establish that a given density is an equilibrium density by showing it has no dissipation. Then we use the Dissipation Theorem to see that if the system is sampled according to this density, this density will be preserved by the specified dynamics. Then thirdly we show that *any* reasonably smooth deviation from this equilibrium density (even under a time reversal mapping: $\mathbf{q}_i \to \mathbf{q}_i, \mathbf{p}_i \to -\mathbf{p}_i, i = 1,N$) produces dissipation and the ensemble average of the time integral of this dissipation (for a time interval of any finite but nonzero duration) is strictly positive. This proves that for the specified dynamics, if the system is T-mixing, the equilibrium distribution (even under the time reversal mapping) is unique among the class of reasonably smooth distributions. This goes beyond what is possible using the constructs of the ergodic theory proofs of relaxation to equilibrium. Next the ΩT-mixing proof shows that the ensemble average of all phase functions that are odd under time reversal approach zero in the infinite time limit. [6] This is because in the long time limit for ΩT-mixing systems, the ensemble average of even, phase functions become stationary. This proves that in the infinite time limit the distribution must itself be even in the momenta. Finally the $\Omega$T-mixing proof shows that in the late time limit, the time integral (over any finite duration) of the ensemble average of the instantaneous dissipation goes to zero. However we already know that in T-mixing systems, the only even, smooth distribution that has a zero ensemble average for the dissipation function, is the equilibrium distribution. We conclude that the distribution that the system is *apparently* approaching is the equilibrium distribution.

---

[3] A ΩT-mixing system is one for which, $\lim_{t \to \infty} \int_0^t \langle \Omega(\Gamma) B(S^t \Gamma) \rangle_0 \, ds = L_A \in \mathbf{R}$ for all phase variables, $B$ and where $L_A$ is a finite real number.

We now return to our generalization of the ergodic theory proof of relaxation to equilibrium. Using our definition of an equilibrium system we write the equilibrium distribution as,

$$f_{eq,h}(\mathbf{\Gamma}) = \frac{\exp[-h(\mathbf{\Gamma})]}{\int_D d\mathbf{\Gamma} \exp[-h(\mathbf{\Gamma})]} \qquad (10)$$

where $h(\mathbf{\Gamma})$ is some real, integrable function of the (possibly augmented) phase space vector, $\mathbf{\Gamma}$ defined over some domain, *D*. *All* equilibrium distributions must be expressible in a form given by (9). Equilibrium distributions must be autonomous because by definition the dissipation must be zero and therefore they must be time independent.

Again this initial distribution (10) will need to be reasonably smooth. We compute the average of some physical phase space function $A(\mathbf{\Gamma})$ at some time *t* with an initial distribution $f_0(\mathbf{\Gamma}) \neq f_{eq,h}(\mathbf{\Gamma})$ and undergoing dynamics that would preserve (10):

$$\langle A(t)\rangle_0 = \int_D d\Gamma\, A(S^t\Gamma) f_0(\Gamma)$$

$$= \frac{\int_D d\Gamma\, A(S^t\Gamma) f_0(\Gamma)\exp[h(\Gamma)]\exp[-h(\Gamma)]}{\int_D d\Gamma\, \exp[-h(\Gamma)]} \int_D d\Gamma\, \exp[-h(\Gamma)]$$

$$= \langle A(S^t\Gamma) f_0(\Gamma)\exp[h(\Gamma)]\rangle_{eq,h} \int_D d\Gamma\, \exp[-h(\Gamma)]$$

$$\xrightarrow[t\to\infty]{} \langle A(\Gamma)\rangle_{eq,h} \langle f_0(\Gamma)\exp[h(\Gamma)]\rangle_{eq,h} \int_D d\Gamma\, \exp[-h(\Gamma)] \qquad (11)$$

$$= \langle A(\Gamma)\rangle_{eq,h} \frac{\int_D d\Gamma\, f_0(\Gamma)\exp[h(\Gamma)]\exp[-h(\Gamma)]}{\int_D d\Gamma\, \exp[-h(\Gamma)]} \int_D d\Gamma\, \exp[-h(\Gamma)]$$

$$= \langle A(\Gamma)\rangle_{eq,h} \int_D d\Gamma\, f_0(\Gamma)$$

$$= \langle A(\Gamma)\rangle_{eq,h}$$

where $\langle\cdots\rangle_{eq,h}$ denotes an average with respect to the mixing equilibrium distribution given by (8). We note that the partition function associated with $f_0(\Gamma)$ and $\int_D d\Gamma\, \exp[-h(\Gamma)]$ must be finite and non-zero otherwise the derivation cannot be completed. We also require a form of ergodic consistency for all $\Gamma$ such that if $f_0(\Gamma) \neq 0$, we require $\exp[-h(\Gamma)], \exp[h(\Gamma)] \neq 0$.

These latter points mean that the derivation cannot be extended to nonequilibrium steady states because the relevant "partition functions" and the relevant functions $\exp[-h(\mathbf{\Gamma})], \exp[h(\mathbf{\Gamma})]$, would be singular. In this respect, we observe that Sinai's proof can be extended to singular mixing distributions, [1] but unlike the cases we usually consider in physics, the starting state must also be represented by a singular distribution. In our notation this system could be treated if $D$ was considered to be the domain of the mixing state for which the density in the ostensible phase space was singular, and $f_0(\mathbf{\Gamma})$ was only defined in this domain. However, for a dissipative steady state this domain would be a very complicated object of zero volume in the ostensible phase space. Therefore, specifying ergodic consistency is a way of realizing that the initial and final distributions are "absolutely continuous" with respect to each other.

As was the case for Hamiltonian dynamics (9) proves the relaxation of averages to their equilibrium values. We cannot know from this proof whether the distribution relaxes towards its equilibrium form or not. This is only established using the $\Omega T$-mixing proof and realising that the averages of physical quantities only depend on the low order terms in the Green's expansion of the distribution function.

The proof of relaxation to equilibrium given above complements our previous proof of relaxation to equilibrium in thermostatted or indeed barostatted systems that are $\Omega T$-mixing. [3, 4, 6] The proof given above is considerably shorter than the proof using $\Omega T$-mixing. However it must be supplemented with our recent definition of an equilibrium system. It also gives no information about relaxation timescales (which are important in the computation of the values of transport coefficients) and it gives no information in itself about the relaxation of distribution functions. It only says that in the infinite time limit averages of physical properties approach their equilibrium values.

The ΩT-mixing proof has much more information in this respect [3,4,6].

We know from our ΩT-mixing equilibrium relaxation theorems that the necessary and sufficient conditions for the relaxation of an initial ensemble to the corresponding equilibrium ensemble is that the system must be ΩT-mixing. Of course in contradistinction, the T-mixing condition is a sufficient but not necessary condition.

The mere existence of a mixing equilibrium state implies that the transient states are ΩT-mixing and that all systems possessing mixing equilibrium states will in fact relax towards these states at long times – at least as can be inferred from the calculation of averages of physical phase functions from initial states characterized by non-singular distributions.

**Acknowledgements**

We would like to thank the Australian Research Council for support of this research. LR thanks the European Research Council, for funding under the European Community's Seventh Framework Programme (FP7/2007-2013)/ERC grant agreement n 202680. The EC is not liable for any use that can be made on the information contained herein.